\documentclass[%
reprint,
 amsmath,amssymb,
 aps,
]{revtex4-2}

\usepackage{graphicx}
\usepackage{dcolumn}
\usepackage{bm}

\usepackage[caption=false]{subfig}
\usepackage{siunitx, soul, xcolor}
\usepackage{adjustbox}
\usepackage{gensymb}

\usepackage{xparse,xcoffins}

\ExplSyntaxOn
\NewCoffin\imagecoffin
\NewCoffin\labelcoffin

\keys_define:nn { miguel/label }
 {
  label   .tl_set:N = \l_miguel_label_tl,
  labelbox .bool_set:N = \l_miguel_label_box_bool,
  labelbox .default:n = true,
  fontsize .tl_set:N = \l_miguel_label_size_tl,
  fontsize .initial:n = \footnotesize,
  pos .choice:,
  pos/nw .code:n = \tl_set:Nn \l_miguel_label_pos_tl { left,up },
  pos/ne .code:n = \tl_set:Nn \l_miguel_label_pos_tl { right,up },
  pos/sw .code:n = \tl_set:Nn \l_miguel_label_pos_tl { left,down },
  pos/se .code:n = \tl_set:Nn \l_miguel_label_pos_tl { right,down },
  pos/n .code:n = \tl_set:Nn \l_miguel_label_pos_tl { hc,up },
  pos/w .code:n = \tl_set:Nn \l_miguel_label_pos_tl { left,vc },
  pos/s .code:n = \tl_set:Nn \l_miguel_label_pos_tl { hc,down },
  pos/e .code:n = \tl_set:Nn \l_miguel_label_pos_tl { right,vc },
  pos .initial:n = nw,
  unknown .code:n   = \clist_put_right:Nx \l_miguel_label_clist
                       { \l_keys_key_tl = \exp_not:n { #1 } }
 }
\clist_new:N \l_miguel_label_clist
\box_new:N \l_miguel_label_box
\box_new:N \l_miguel_label_image_box

\NewDocumentCommand{\xincludegraphics}{O{}m}
 {
  \group_begin:
  \tl_clear:N \l_miguel_label_tl
  \clist_clear:N \l_miguel_label_clist
  \keys_set:nn { miguel/label } { #1 }
  \tl_if_empty:NTF \l_miguel_label_tl
   {
    \miguel_includegraphics:Vn \l_miguel_label_clist { #2 }
   }
   {
    \SetHorizontalCoffin\imagecoffin
     {
      \miguel_includegraphics:Vn \l_miguel_label_clist { #2 }
     }
    \SetHorizontalCoffin\labelcoffin
     {
      \raisebox{\depth}
       {
        \bool_if:NTF \l_miguel_label_box_bool
         { \fcolorbox{white}{white}{\l_miguel_label_size_tl\l_miguel_label_tl} }
         { \l_miguel_label_size_tl\l_miguel_label_tl }
       }
     }
    \SetVerticalPole\imagecoffin{left}{-0pt+\CoffinWidth\labelcoffin/2}
    \SetVerticalPole\imagecoffin{right}{\Width-3pt-\CoffinWidth\labelcoffin/2}
    \SetHorizontalPole\imagecoffin{up}{\Height+10pt-\CoffinHeight\labelcoffin/2}
    \SetHorizontalPole\imagecoffin{down}{3pt+\CoffinHeight\labelcoffin/2}
    \use:x{\JoinCoffins\imagecoffin[\l_miguel_label_pos_tl]\labelcoffin[vc,hc]} 
    \TypesetCoffin\imagecoffin
   }
   \group_end:
 }
\NewDocumentCommand{\setlabel}{m}
 {
  \keys_set:nn { miguel/label } { #1 }
 }

\cs_new_protected:Nn \miguel_includegraphics:nn
 {
  \includegraphics[#1]{#2}
 }
\cs_generate_variant:Nn \miguel_includegraphics:nn { V }

\ExplSyntaxOff

\begin{document}
\preprint{APS/123-QED}

\title{Interface transparency to orbital current}

\author{Igor Lyalin}
\affiliation{Department of Physics, The Ohio State University, Columbus, Ohio 43210, USA}
\author{Roland K. Kawakami}
\email{kawakami.15@osu.edu}
\affiliation{Department of Physics, The Ohio State University, Columbus, Ohio 43210, USA}

\begin{abstract}
The transport of spin currents across interfaces is relatively well studied, while the transport properties of orbital currents are just starting to be examined.
In Cr/Ni bilayers, the spin-orbit torque (SOT) due to the orbital current generated in the Cr layer is believed to dominate over torques of other origins.
In this work, we study SOT in Cr/X/Ni trilayers, where X is an ultra-thin spacer of a different material.
Using the SOT as a proxy for the orbital current transferred from the Cr to the Ni layer, we compare Cr/X/Ni results to Pt/X/Ni, the system in which spin current generated in the Pt layer plays a dominant role.
We find that across 12 different spacers the apparent interface transparency to the orbital current is comparable or larger than to the spin current. 

\end{abstract}

\flushbottom
\maketitle
\thispagestyle{empty}

\section{Introduction}

The orbital Hall effect (OHE) describes a phenomenon in which an electrical charge current generates a transverse flow of orbital angular momentum.
It is predicted theoretically that the OHE can exist in solids even in the absence of spin-orbit coupling (SOC) and give rise to the spin Hall effect (SHE) when SOC is present~\cite{tanaka_intrinsic_2008,kontani_giant_2009, go_intrinsic_2018,jo_gigantic_2018,bhowal_intrinsic_2020}. 
Thus, one can argue that the OHE is more fundamental than SHE~\cite{kontani_giant_2009,go_intrinsic_2018}.
Recently, there has been a growing amount of experimental observations supporting the idea of the orbital Hall effect.
Unexpectedly large spin-orbit-torque (SOT) signals has been measured in seemingly trivial systems without heavy metals, e.g.\ CuO/FM, Cr/FM, and attributed to the orbital currents generated in non-magnetic layers~\cite{ding_harnessing_2020,kim_nontrivial_2021,lee_efficient_2021,sala_giant_2022,hayashi_observation_2023}.
It has also been observed that while SOT is usually large and negative in Ta/FM bilayers, it depends on the choice of FM and is positive in Ta/Ni~\cite{lee_orbital_2021}.
This was attributed to the torque due to the OHE dominating over the SHE-induced torque.
More recently, an orbital accumulation due to the OHE on the surface of a single layer of light metals Ti~\cite{choi_observation_2023} and Cr~\cite{lyalin_magneto-optical_2023} was detected by magneto-optic Kerr effect, confirming the existence of the OHE more directly.

While spin transport is relatively well studied, the transport properties of orbital currents have just started to be investigated.
Interface transparency to orbital current is an important question from both, fundamental and application, points of view, since to utilize orbital currents for spintronic applications the efficient injection across NM/FM is required.
It has been suggested theoretically that orbital transport being dependent on the orbital hybridization at the interface is more sensitive to interface quality and disorder than spin transport~\cite{go_orbital_2020}.

In this work, we study interface transparency to orbital current in Cr/X/Ni trilayers, where X is an ultra-thin spacer of a different material.
In Cr/Ni bilayers, the SOT due the orbital current generated in Cr layer is believed to dominate over torques of the other origin~\cite{lee_efficient_2021,sala_giant_2022}.
We use the experimentally measured SOT as a proxy for the orbital current transferred from Cr to Ni layer.
Comparing Cr/X/Ni to a trilayer system in which spin current plays the major role, Pt/X/Ni, we find that across 12 different spacers the apparent interface transparency to the orbital current is comparable or larger than to the spin current.

\section{Experiment}

Using current-modulated magneto-optic Kerr effect (MOKE) technique~\cite{fan_quantifying_2014}, we measure the damping-like spin-orbit torque in Cr/X/Ni (20/0.8/5\,nm) and Pt/X/Ni (20/0.8/5\,nm) trilayers, where the numbers in parentheses represent the thickness in nanometers.
We study 15 different spacer layers X: Al, Ti, Fe, Co, Cu, Ge, Pd, Ag, Sn, Ta, W, Pt, Au, NiO, MgO.
The films are deposited at room temperature on Mg(100) substrates by a combination of electron beam evaporation and magnetron sputtering in a chamber with a base pressure
of $2 \times 10^{-7}$\,Torr. 
To protect the films from oxidation, they are capped with MgO/Cr (3/3\,nm).
An example of the full stack sequence is MgO/Cr/Cu/Ni/MgO/Cr (substrate/20/0.8/5/3/3\,nm).
For Pt/X/Ni samples a 1\,nm thick Cr seed layer is used to improve adhesion between a Pt layer and an MgO substrate.
During the growth, the thickness of the deposited materials is monitored by a quartz crystal, which readings are calibrated by X-ray reflectivity measurements.
We estimate the error of the spacer layer thickness to be about $20\%$.

The films are patterned into 20\,$\mu$m wide devices by photolithography and argon ion milling.
Measuring the resistivity of Cr/MgO/Cr (1/3/3\,nm) stack, we estimate that less than 1\% of total current applied to the devices is flowing through the seed layer and partially oxidized Cr cap.
The average resistivities of the entire Cr/X/Ni and Pt/X/Ni stacks are $(96\pm19)\,\mu\Omega$cm and $(31\pm3)\,\mu\Omega$cm, respectively.
We do not attempt to estimate the resistivities of the individual layers since the resistivity of thin multilayers depends strongly on interfacial scattering, while the growth, and as a result, the resistivity of Ni layer, is affected by the choice of the underlying spacer X.

\begin{figure}[h!]    
\includegraphics[width=0.32\textwidth]{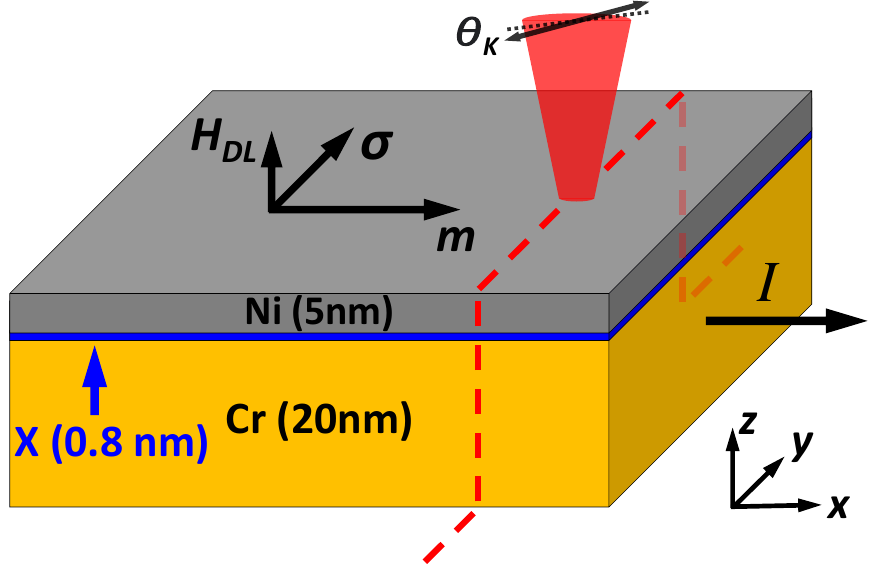}\vspace{-0.2cm}\hfill
  \caption{
  Schematics of the current-modulated polar MOKE experiment.
  A laser is scanned across a Cr/X/Ni device.
  The Kerr rotation detects the current-induced out-of-plane component of magnetization due to the damping-like effective SOT field $\boldsymbol{H_{DL}} \sim [ \boldsymbol{m} \times \boldsymbol{\sigma} ]$ and the out-of-plane Oersted field.   
  }
\label{fig:current_modulated_moke_schematics}
\end{figure}

A mode-locked Ti:Sapphire laser tuned to $\lambda = 800$\,nm wavelength is utilized for the polar MOKE measurements.
The sample is mounted on a motorized XY stage for scanning in the sample plane. 
To exclude contributions from quadratic MOKE, the light is linearly polarized at $45\degree$ relative to the applied magnetic field~\cite{fan_all-optical_2016}.
The beam is focused on the sample to a spot size of about 2\,$\mu$m.
The laser power incident on the sample is $800\,\mu$W.
The Kerr rotation of the reflected light is measured by a combination of a half wave plate, Wollaston prism, and balanced photodetector.
The signal from the balanced photodetector is passed to a lock-in amplifier, which measures the current-induced Kerr rotation by demodulating at the first harmonic of the applied sinusoidal current of 41227\,Hz frequency.
The schematic of the experimental geometry is shown in Fig.~\ref{fig:current_modulated_moke_schematics}.

\section{Results and Discussion}

\begin{figure}[h!]    
\subfloat{\xincludegraphics[width=0.24\textwidth,label=(a)]{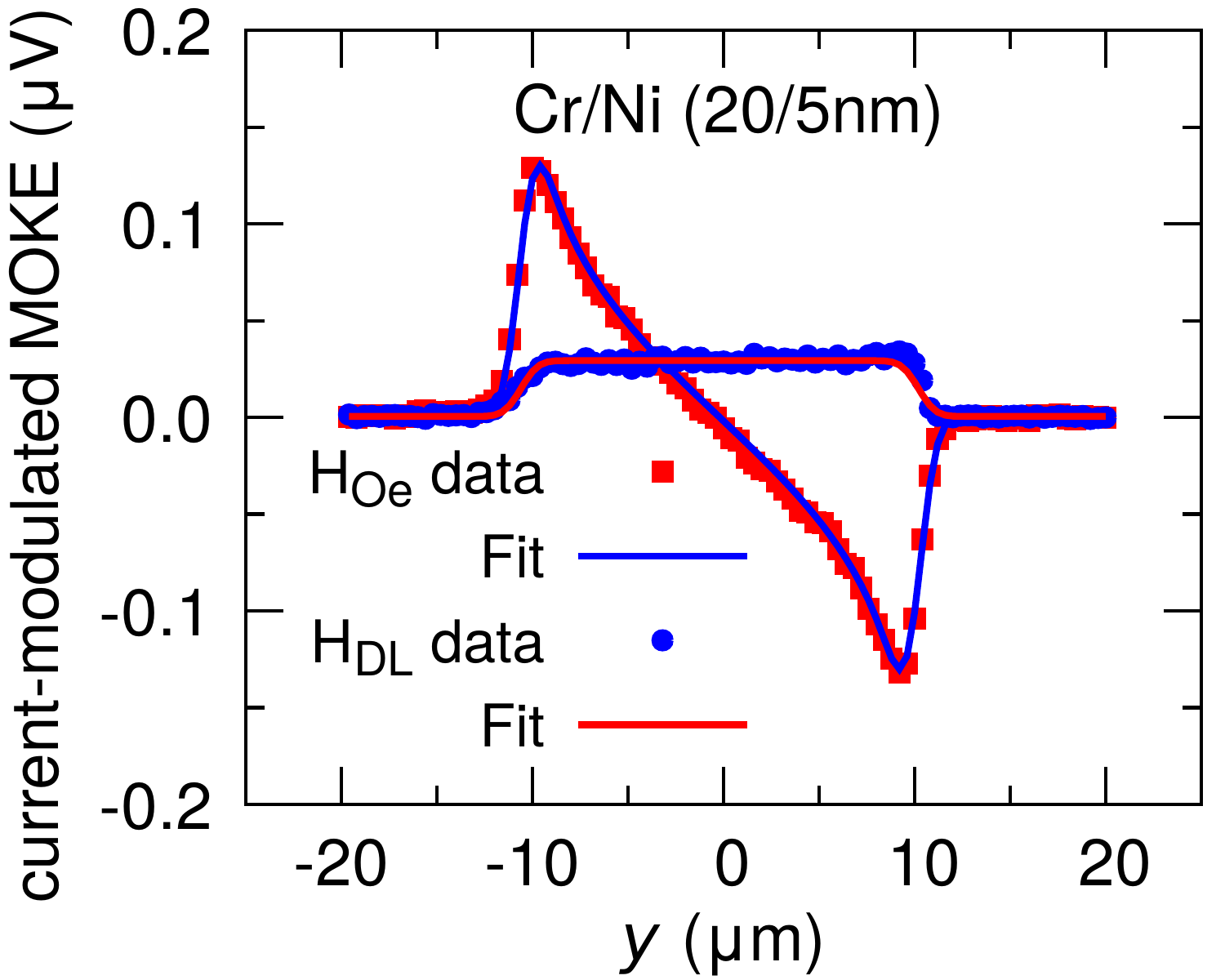}\label{fig:CrNi_linecut}}\hfill
\subfloat{\xincludegraphics[width=0.24\textwidth,label=(b)]{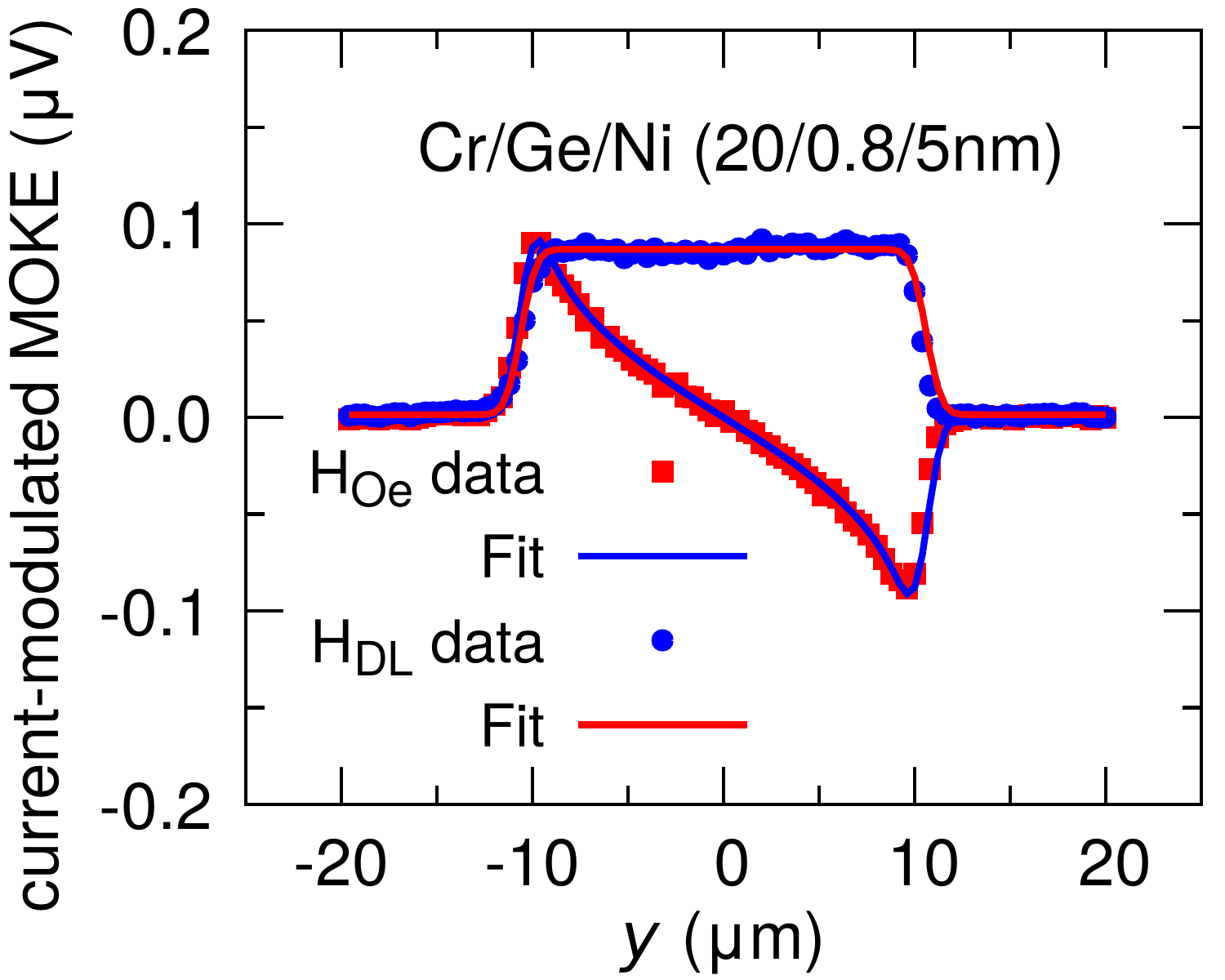}\label{fig:CrGeNi_linecut}}\vspace*{-0.4cm}\hfill
\subfloat{\xincludegraphics[width=0.24\textwidth,label=(c)]{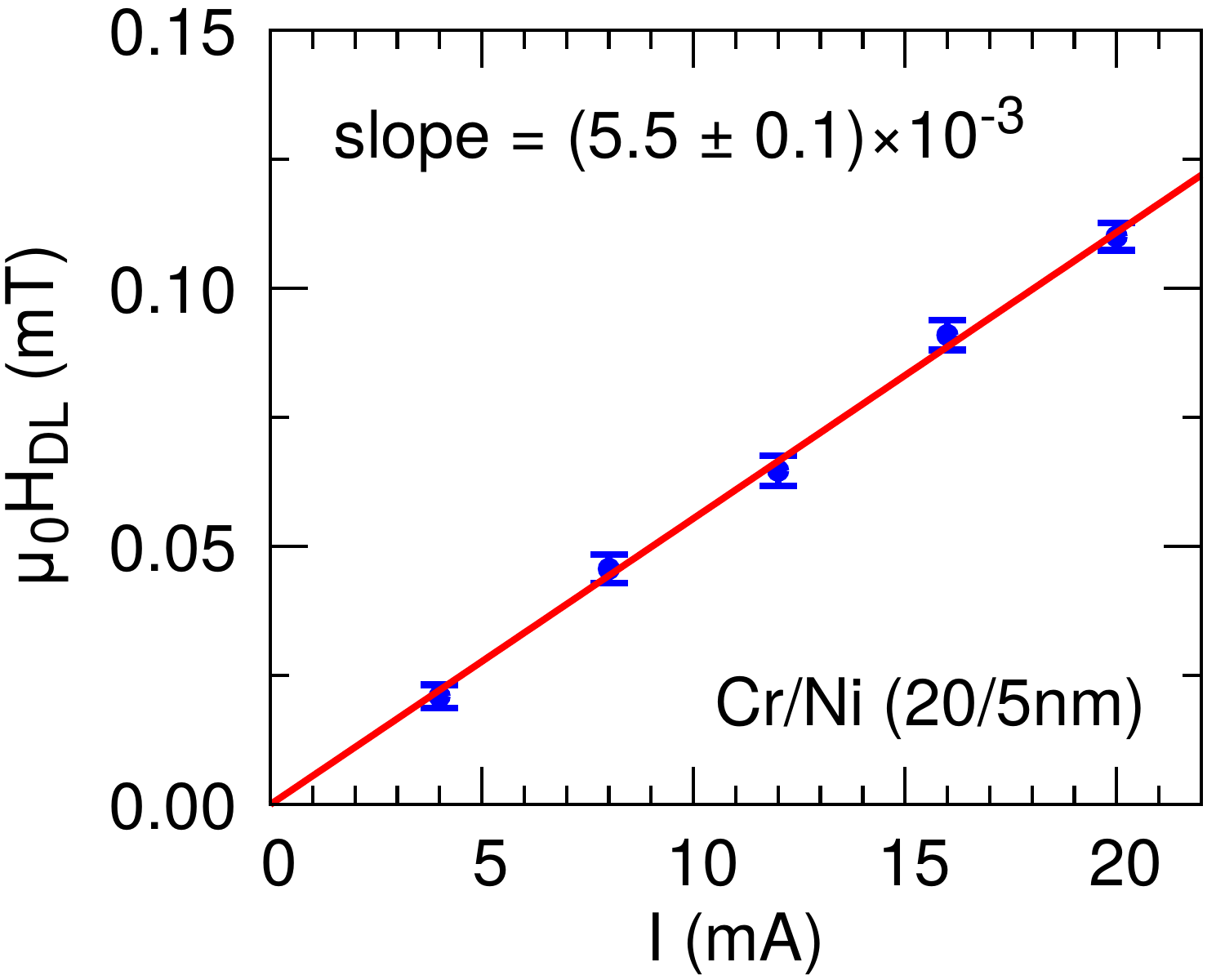}\label{fig:CrNi_H_DL_vs_current}}\hfill
\subfloat{\xincludegraphics[width=0.24\textwidth,label=(d)]{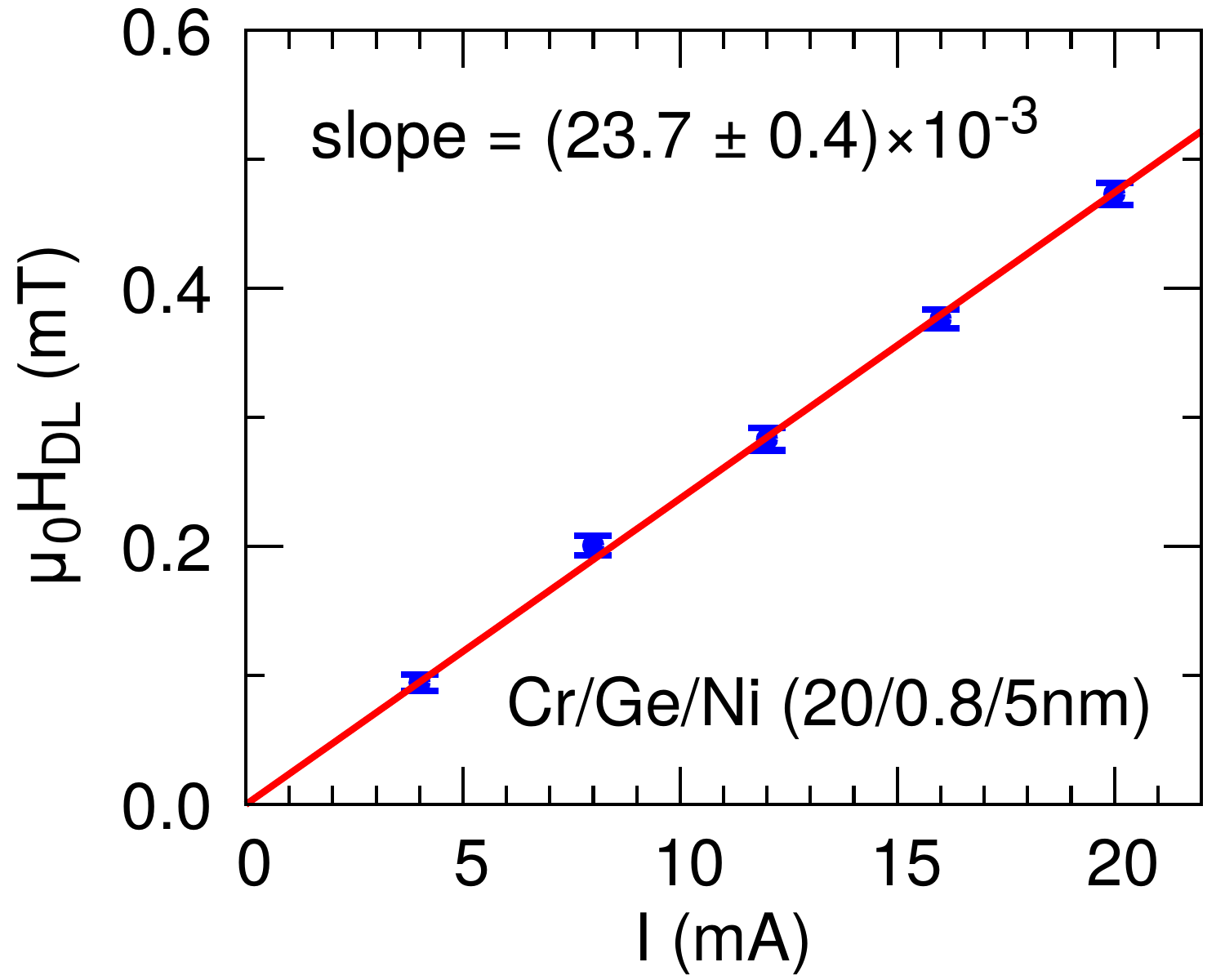}\label{fig:CrGeNi_H_DL_vs_current}}\vspace*{-0.2cm}\hfill
  \caption{
  Current-modulated MOKE line scan across a 20\,$\mu$m wide channel of (a) Cr/Ni (20/5\,nm) and 
  (b) Cr/Ge/Ni (20/0.8/5\,nm) at current $I=20$\,mA, corresponding to current density $j=4 \times 10^{10}$\,A/m$^2$.
  Effective damping-like SOT field, $\mu_0 H_{DL}$, as a function of applied current for (c) Cr/Ni and (d) Cr/Ge/Ni. 
  }
  \label{fig:MOKE_Cr_main}
\end{figure}

In Fig.~\ref{fig:CrNi_linecut}, we show the current-modulated MOKE data measured for Cr/Ni.
Measuring line scans across the 20\,$\mu$m wide device for the Ni magnetization parallel or antiparallel to the applied current, we separate the MOKE data into the symmetric part (blue dots in Fig.~\ref{fig:CrNi_linecut}), and antisymmetric part (red squares in Fig.~\ref{fig:CrNi_linecut}).
The symmetric component is proportional to the effective damping-like SOT field $\sim$$H_{DL}$, while antisymmetric component is proportional to the out-of-plane Oersted field $\sim$$H^{oop}_{Oe}$~\cite{fan_quantifying_2014}.
Using calibration by the Oersted field, the DL-SOT signal is converted from MOKE voltages to actual magnetic field values.
Effective DL-SOT field as a function of applied current $I$ for the Cr/Ni device is plotted in Fig.~\ref{fig:CrNi_H_DL_vs_current}. 
We observe that the signal scales linearly with the current, as expected for the DL-SOT.
The slope of the current dependence, $\mu_0 H_{DL}/I$, dimensions of the device, and the resistivity of the whole stack $\rho_{xx}$ is used to determine the damping-like SOT efficiency per applied electric field~\cite{nguyen_spin_2016}:
\begin{equation}
    \xi_E = \dfrac{2e}{\hbar} M_s t_{FM} \dfrac{\mu_0 H_{DL}}{E}\,,
    \label{eq:xi_E}
\end{equation}
where $e$ is the electron charge, $\hbar$ is the reduced Planck constant, $t_{FM}$ is the thickness of the Ni layer, $M_s$ is its saturation magnetization, $E$ is the applied electric field.
The surface magnetization $M_st_{FM}$ is measured by superconducting quantum interference device (SQUID) magnetometry.
We note, that we chose $t_{NM}$, the thickness of NM layer, to be larger than the orbital (spin) diffusion length $\lambda^{Cr}_o \approx 6$\,nm~\cite{lee_efficient_2021,lyalin_magneto-optical_2023} ($\lambda^{Pt}_s \approx$\,1--11\,nm~\cite{nguyen_spin_2016,stamm_magneto-optical_2017}).
Since DL-SOT efficiency saturates with the thickness of the NM layer as $\sim [1-\text{sech}(t_{NM}/\lambda^{NM}_{o(s)}]$, for $t_{NM} = 20$\,nm it is close to its saturated value.

Measuring three Cr/Ni samples grown at different times we find that
$\xi_E = (0.17 \pm 0.05)\times10^3\,\Omega^{-1}$cm$^{-1}$, consistent with previous reports~\cite{lee_efficient_2021, sala_giant_2022}.
The positive sign of the torque agrees with the picture of the orbital torque dominating over the negative spin torque generated by the SHE in Cr.
Performing measurements on Pt/Ni we obtain an efficiency $\xi_E = (2.43 \pm 0.11)\times10^3\,\Omega^{-1}$cm$^{-1}$ of the same sign.

\begin{figure*}[t!]
\subfloat{\xincludegraphics[width=1.0\textwidth,label=]{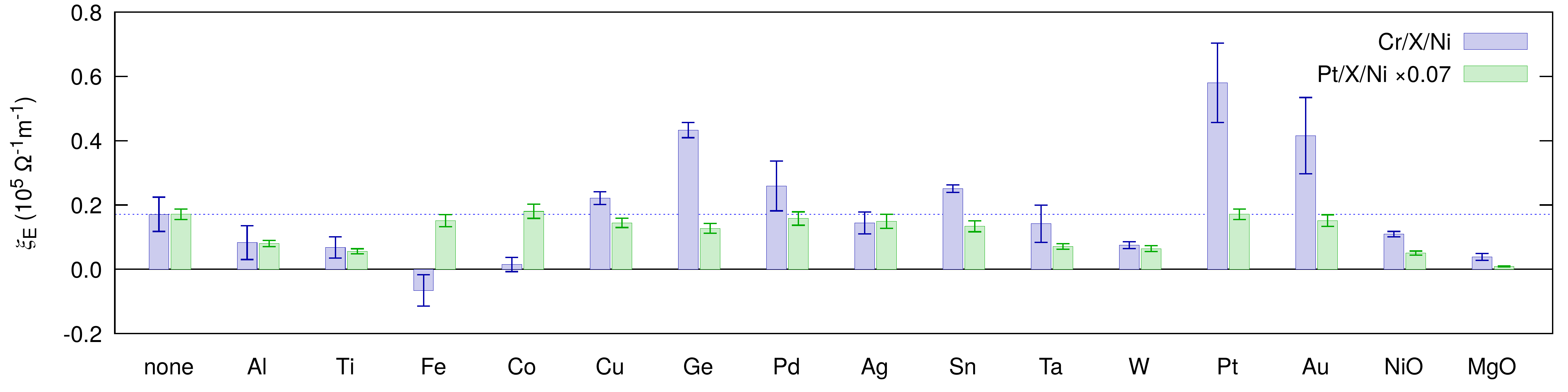}\label{fig:}}\hfill
\caption{
Damping-like SOT efficiency per electric field for 
Cr/X/Ni and Pt/X/Ni trilayers.
The results are plotted side-by-side for each X to compare the behavior of the system dominated by orbital current, Cr/X/Ni, with the system dominated by the spin current, Pt/X/Ni.
The DL-SOT efficiency for all Pt/X/Ni trilayer is scaled by 0.07 to match the Pt/Ni value to the Cr/Ni value.
}  \label{fig:Histogram_xi_E}
\end{figure*}

Figure~\ref{fig:CrGeNi_linecut} and~\ref{fig:CrGeNi_H_DL_vs_current} shows the current-modulated MOKE data for Cr/X/Ni with one of the spacers, X = Ge.
A histogram in Figure~\ref{fig:Histogram_xi_E} summarizes DL-SOT efficiency results for all spacers.
Cr/X/Ni and Pt/X/Ni results are plotted side-by-side for each X to compare the behavior of the system dominated by orbital current with the system dominated by the spin current.
Each Cr/X/Ni column is the average of two samples grown at different times. 
In the following, we discuss the results summarized in the Figure~\ref{fig:Histogram_xi_E}, making a number of observations.

\subsection{Non-magnetic metal spacers}
Across 11 non-magnetic spacer materials (Al, Ti, Cu, Ge, Pd, Ag, Sn, Ta, W, Pt, Au) apparent interface transparency to orbital current is comparable or larger than to spin current.
We define the apparent orbital interface transparency ratio as $\mathcal{T}_o^{\rm X} = \frac{\xi^{\rm {Cr/X/Ni}}_E}{\xi^{\rm {Cr/Ni}}_E}$ and the apparent spin interface transparency ratio as $\mathcal{T}_s^{\rm X} = \frac{\xi^{\rm {Pt/X/Ni}}_E}{\xi^{\rm {Pt/Ni}}_E}$.

\begin{figure}[h!]    
\includegraphics[width=0.2\textwidth]{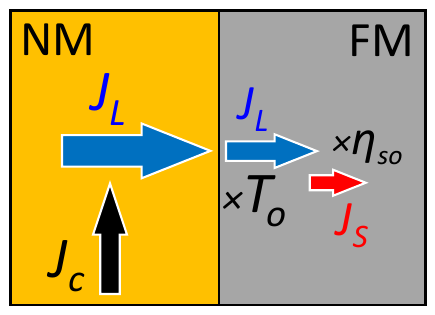}\vspace{-0.2cm}\hfill
  \caption{
  Schematics of the orbital current transferred across an NM/FM interface. 
  The orbital current generated by a NM layer is reduced by a factor $T_o$ at the interface, before being converted into a spin current by the spin-orbit coupling in the FM layer, with the conversion ratio $\eta_{so}^{FM}$.
  The converted spin current exerts an orbital torque on the FM magnetization. 
  }
  \label{fig:orbital_transparency_schematics}
\end{figure}

\textit{Light metals and interface transparencies.} The DL-SOT efficiency due to the SHE can be expressed as $\xi_E = T_s \sigma^{\rm NM}_{\rm SH}$, where $T_s$ is the spin interface transparency, $\sigma^{\rm NM}_{\rm SH}$ is spin Hall conductivity of the NM layer.
According to the current understanding~\cite{go_orbital_2020}, the orbital torque is a result of the orbital current generated by the OHE in the NM layer, transferred across NM/FM interface, and converted to the spin current by the SOC of the FM layer.
The resultant spin current exerts a torque on the FM magnetization, as the orbital current cannot interact with the magnetization directly~\cite{go_orbital_2020}.
Thus, the DL-SOT efficiency due to the OHE can be expressed as $\xi_E = T_o \eta^{\rm FM}_{so} \sigma^{\rm NM}_{\rm OH}$, where $T_o$ is the orbital interface transparency, $\eta^{\rm FM}_{so}$ is the conversion ratio from the orbital to spin in the FM layer, proportional to SOC in the FM, $\sigma^{\rm NM}_{\rm OH}$ is orbital Hall conductivity of the NM layer, see schematics in Fig.~\ref{fig:orbital_transparency_schematics}. 
For light 3d transition metals Ti and Cu with weak SOC, it would be reasonable to assume that within the ultra-thin 0.8\,nm spacer a negligible amount of orbital and spin current is generated or destroyed since their orbital and spin diffusion lengths are much larger than the thickness of the spacer~\cite{choi_observation_2023, kim_nontrivial_2021,bass_spin-diffusion_2007}, thus $\sigma^{\rm NM}_{\rm OH}$ should not be modified.
It is also logical to assume that, having weak SOC, Ti and Cu do not significantly modify $\eta^{\rm FM}_{so}$ of Ni, as a spacer with strong SOC, for instance, Pt, does.
Therefore, in a simplistic model of two perfect interfaces, the apparent transparency for X = Ti can be written as $\frac{\xi_E^{\rm Cr/Ti/Ni}}{\xi_E^{\rm Cr/Ni}} = \frac{T_o^{\rm Cr/Ti} T_o^{\rm Ti/Ni}}{T_o^{\rm Cr/Ni}}$.
Under the assumption that the three interfaces between the three similar $3d$ transition metals have the same transparency, we estimate the order of magnitude of orbital transparency as $T_o^{\rm Cr/Ni} \approx T_o^{\rm Cr/Ti} \approx T_o^{\rm Ti/Ni}\approx40\%$. 
The same logic applied to X = Cu yields an unrealistic value of $\approx 130\%$.
For comparison, the spin interface transparency $T_s$ of Py/Pt and Py/Ni/Pt interfaces (Py = Ni$_{80}$Fe$_{20}$) was investigated in Ref.~\cite{zhang_role_2015}, and estimated to be $\approx 25\%$. 
These estimates of orbital interface transparency for X = Ti, Cu, are comparable to or larger than the spin interface transparency of Py/Pt and Py/Ni/Pt.
We emphasize that we do not attempt to give a precise quantitative evaluation of the orbital interface transparency by this oversimplified reasoning, but rather to perform a qualitative analysis. 
For discussion later in the paper, we take the lower estimate of $\approx 40\%$ for the Cr/Ni orbital interface transparency $T_o^{\rm Cr/Ni}$.
Lastly, we note that the model above implies that the two interfaces can be treated independently which might not be a good assumption for a very thin spacer.
We also observe that $\mathcal{T}_o^{\rm X} \gtrsim \mathcal{T}_s^{\rm X}$ for the other light elements X = Al and Ge.

Based on these observations for light elements with weak SOC, we tentatively conclude that not only the apparent orbital interface transparency $\mathcal{T}_o^{\rm X}$ (which could include SOC effects in the spacer or FM as discussed later), but also the actual orbital interface transparency could be comparable or larger than the spin interface transparency.

\textit{Orbital-to-spin conversion in the Ni layer.} Using $\xi_E = T_o \eta^{\rm FM}_{so} \sigma^{\rm NM}_{\rm OH}$, theoretically calculated
$\sigma^{\rm Cr}_{\rm OH} = 7\times10^5\,\Omega^{-1}\rm m^{-1}$~\cite{salemi_first-principles_2022}, and experimentally measured $\xi_E = (0.17 \pm 0.05)\times10^5\,\Omega^{-1}\rm m^{-1}$ we can estimate $T^{\rm Cr/Ni}_o \eta^{\rm Ni}_{so}$ to be $\approx 2.4\%$. 
Using the estimate of $40\%$ for $T^{\rm Cr/Ni}_o$ translates into $\eta^{\rm Ni}_{so} \approx 6\%$.
Therefore, while increasing interface transparency is important for utilizing orbital torque in SOT-based applications, increasing the orbital-to-spin conversion in FM layer might be a more promising route, as there is more room for improvement.
For example, this can be achieved by alloying Ni and a HM with a large and positive SOC, e.g. using Ni$_{90}$Pt$_{10}$ as a ferromagnetic layer. 

\textit{Orbital character of the spacer (p vs.~d).} We observe that the apparent orbital transparency for Cu and Ge spacers is larger than their apparent spin transparency.
The SOC strength increases with atomic number $Z$, thus one could try to explain the monotonic increase Ti $\rightarrow$ Cu $\rightarrow$ Ge of the $\mathcal{T}_o^{X}$, by enhanced $\eta^{\rm Ni}_{so}$ due to additional orbital-to-spin conversion in the spacer layer.
However, the $\mathcal{T}_o^{Cu} > \mathcal{T}_o^{Ag}$ and $\mathcal{T}_o^{Ge} > \mathcal{T}_o^{Sn}$, of the heavier elements in the same group in the periodic table with even larger SOC strength.
Thus, the SOC of these spacers alone cannot account for the enhanced orbital torque.
Instead, we attribute it to the enhanced interface transparency by Cu and Ge spacers.
The 2.5-fold increase of the interface transparency for X = Ge (compared to without a spacer) might be due to $4p$-orbitals of Ge, indicating that $p$-$d$ hybridization at the interface facilitates orbital transport in comparison with the case of Cr/Ni interface that have only $s$- and $d$-orbitals.
If $T^{\rm Cr/Ni}_o \approx 40\%$, then Cr/Ge/Ni interface is close to being fully transparent.
Thus, to maximize the orbital interface transparency, and, as a result, orbital torque efficiency, elements with different orbital character on the opposite sides of NM/FM bilayers might be preferable.
We note, that so far, the orbital torque has been studied mostly in $3d/3d$, $4d/3d$, and $5d/3d$ bilayers.  

 \textit{Heavy metals with positive SOC.} The insertion of $4d$ and $5d$ elements with positive SOC (Pd, Ag, Sn, Pt, Au) seems to preserve or increase the $\mathcal{T}_o^{\rm X}$.
We attribute this to the increase of the effective orbital-to-spin conversion parameter $\eta^{\rm Ni}_{so}$ in the Ni layer, by being adjacent to a material with a larger positive SOC strength. 
Consistent with this picture,  $\mathcal{T}_o^{\rm X}$ is larger for the heavier $5d$ spacers (Pt and Au) than for $4d$ elements (Pd, Ag, Sn).
Pt and Pd are known to be strong spin sources, so the charge current even in 0.8\,nm-thick layers can generate a positive spin current that can enhance the measured torque.
This could explain why $\mathcal{T}_o^{\rm Pd} \gtrsim \mathcal{T}_o^{\rm Ag}$ and $\mathcal{T}_o^{\rm Pt} \gtrsim \mathcal{T}_o^{\rm Au}$.
For X = Sn, both, enhanced $\eta^{\rm Ni}_{so}$ due to the Sn SOC and enhanced $T_{o}$ due to $5p$-$3d$ hybridization, can play a role. 

\textit{Heavy metals with negative SOC.} The behavior of the apparent transparency for Ta and W is consistent with the interpretation given in the previous paragraph.
Having a negative SOC, Ta, and W decrease the positive value of $\eta^{\rm Ni}_{so}$. As a result, the measured DL-SOT torque is reduced.
The spacer thickness is likely not large enough to fully convert the positive orbital current into a negative spin current.
Furthermore, it is not obvious that for a larger thickness of Ta (W), the torque would change the sign, as for thicker Ta layers the positive orbital torque due to the OHE in Ta has been shown to dominate over the negative SHE-induced torque in Ta/Ni bilayers~\cite{lee_orbital_2021}.  

\subsection{Ferromagnetic metal spacers}
The only spacers that decrease orbital torque in Cr/X/Ni to zero, or even reverse its sign, are magnetic spacers, Fe and Co. 
We attribute this to the $\eta^{\rm FM}_{so}$ being negative in Fe, $\eta^{\rm Fe}_{so} < 0$, and almost zero in Co, $\eta^{\rm Co}_{so} \approx 0$.
This agrees with the experimental observation of the self-induced anomalous SOT (ASOT) in single layers of Fe, Co and Ni~\cite{wang_anomalous_2019}.
The ASOT arises from the SOC of FM layer, that generates the spin current that acts on its own magnetization.
In the reference~\cite{wang_anomalous_2019}, the anomalous SOT was found to be negative in Fe, a small positive in Co, and the largest positive in Ni, although this contradicts theoretical calculations in ref.~\cite{lee_orbital_2021}, where the spin-orbit correlation at the Fermi level were calculated to be positive for all the three ferromagnetic materials.
We observe that the DL-SOT is negative in Cr/Fe/Ni and half of the value of Cr/Ni.
The DL-SOT in Cr/Co/Ni is close to zero.
If the orbital diffusion length in Fe and Co would be much larger than 0.8\,nm, then assuming similar to Cr/Ni orbital transparency at the interface, the orbital current would reach Ni layer and be converted to the positive spin current, thus generating a positive DL-SOT.
However, we observe the opposite.
Two possible scenarios can explain the reversal of the torque for the X = Fe: a) the orbital diffusion length in Fe is ultra-short, comparable with the thickness of the spacer $\sim$\,$0.8$\,nm, which leads an efficient orbital-to-spin conversion before the orbital current reaches the Ni layer, b) the orbital transparency of the Cr/Fe interface is much smaller than $T_{o}^{\rm Cr/Ni}$ which leads to the negative spin torque due to the SHE in Cr dominating the measurement.
The observation of the apparent orbital transparency being comparable or higher than spin transparency for all other spacers makes us to believe that the scenario a) is more likely.
This scenario can also explain the nearly zero DL-SOT in Cr/Co/Ni.
The small positive $\eta^{\rm Co}_{so}$ and ultra-short $\lambda^{\rm Co}_{o}$ results in a smaller positive orbital torque than in Cr/Ni which is compensated by the negative SHE-induced torque.
We note, that a very short $\lambda_{o}$ in FMs agrees with earlier theoretical calculations~\cite{go_orbital_2020}, but contradicts to later calculations~\cite{go_long-range_2023}.

\subsection{Antiferromagnetic insulator spacers}
The spin current from heavy metals can interact with magnons in magnetic insulators.
It has been shown that a thin layer of NiO can transfer spin current from HM to FM for temperatures above and below the N\'eel temperature ~\cite{wang_magnetization_2019,zhu_fully_2021,zhu_maximizing_2021,zhu_sign_2022}.
Using NiO as a spacer, we test whether the orbital current can be transferred by magnons as well.
Unlike other spacers, the 0.8\,nm of NiO is synthesized by oxidation.
We reverse the stack sequence, first growing 5.5\,nm of Ni on MgO substrate and then oxidizing the top 0.5\,nm of Ni in 1 mTorr of oxygen atmosphere for 1\,minute to the self-limiting value of  0.8\,nm of NiO.
This oxygen exposure is more than 100 times of the minimal oxygen dose required~\cite{holloway_chemisorption_1981,mitchell_kinetic_1982,lambers_room-temerature_1996}.
After that we deposit 20\,nm of Cr on the top and perform current modulated MOKE measurement through the MgO substrate side. 
We find that the apparent transparency of NiO is about 60\% which is larger than the value for the Ti and Al spacers.
A NiO spacer grown in the same way for Pt/NiO/Ni reduces DL-SOT efficiency to the 30\% of the Pt/Ni value.
Thus, the apparent orbital interface transparency of NiO is larger than its spin transparency. 
In addition, we perform measurements on Cr/MgO/Ni as a control sample.
We observe that DL-SOT is 3 times smaller with the MgO, than with the NiO spacer.
This indicates that electron tunneling across a 0.8\,nm insulating layer cannot account for the large observed $\mathcal{T}_o^{\rm NiO}$ and a
magnon-mediated mechanism of transferring orbital angular momentum from Cr to Ni probably plays a role. 
We attribute the non-zero torque in Cr/MgO/Ni trilayers to the presence of pinholes in the MgO layer.
We note, that NiO grown by oxidation is pinhole free~\cite{holloway_chemisorption_1981,kitakatsu_surface_1998}.
A detailed study is needed to investigate the exact mechanism by which conduction electrons in a metal can transfer their orbital angular momentum to magnons in a magnetic insulator.
Recent theoretical works have investigated the orbital angular momentum of magnons~\cite{neumann_orbital_2020,fishman_orbital_2023}, furthermore, an intrinsic magnon OHE was proposed~\cite{go_intrinsic_2023}.

\section{Conclusions}

In this work, we studied the spin-orbit torque efficiency in Cr/X/Ni trilayers with different ultra-thin spacers.
Using the SOT as a proxy for the orbital current transferred from Cr to Ni layer, we compared Cr/X/Ni to Pt/X/Ni, the system in which spin current plays a dominant role.
The interpretation of the experimental results relies on an assumption that the positive SOT in Cr/Ni is due to the orbital torque dominating over the torques of other origin~\cite{lee_efficient_2021,sala_giant_2022}: spin torque, interfacial torque, and anomalous torque.
Contrary to the theoretical suggestion that the orbital current is more sensitive to interface quality and disorder than the spin current~\cite{go_orbital_2020}, we found that across 12 different spacers, the apparent orbital interface transparency is comparable to or larger than the spin interface transparency. 
In addition to this principal result, the most interesting spacers are: a) Ge, which increases the apparent orbital interface transparency by a factor of 2.5, b) ferromagnetic Fe and Co, which suppress or even reverse the sign of the spin-orbit torque, and c) insulating NiO, capable of a magnon-mediated transfer of orbital angular momentum. 

Our work provides a broad materials perspective on the orbital transport across interfaces and can serve as a starting point for future in-depth studies focused on a specific material combination, thicknesses, and growth optimization.

\section*{Acknowledgements}

We thank Peter M. Oppeneer for insightful discussions.
{This research was primarily supported by} the Center for Emergent Materials, an NSF MRSEC, under award number DMR-2011876.
I.L.~acknowledges support from the Ohio State University Presidential Fellowship.

\bibliography{ohe_vs_she_interface.bib}

\begin{thebibliography}{34}%
\makeatletter
\providecommand \@ifxundefined [1]{%
 \@ifx{#1\undefined}
}%
\providecommand \@ifnum [1]{%
 \ifnum #1\expandafter \@firstoftwo
 \else \expandafter \@secondoftwo
 \fi
}%
\providecommand \@ifx [1]{%
 \ifx #1\expandafter \@firstoftwo
 \else \expandafter \@secondoftwo
 \fi
}%
\providecommand \natexlab [1]{#1}%
\providecommand \enquote  [1]{``#1''}%
\providecommand \bibnamefont  [1]{#1}%
\providecommand \bibfnamefont [1]{#1}%
\providecommand \citenamefont [1]{#1}%
\providecommand \href@noop [0]{\@secondoftwo}%
\providecommand \href [0]{\begingroup \@sanitize@url \@href}%
\providecommand \@href[1]{\@@startlink{#1}\@@href}%
\providecommand \@@href[1]{\endgroup#1\@@endlink}%
\providecommand \@sanitize@url [0]{\catcode `\\12\catcode `\$12\catcode `\&12\catcode `\#12\catcode `\^12\catcode `\_12\catcode `\%12\relax}%
\providecommand \@@startlink[1]{}%
\providecommand \@@endlink[0]{}%
\providecommand \url  [0]{\begingroup\@sanitize@url \@url }%
\providecommand \@url [1]{\endgroup\@href {#1}{\urlprefix }}%
\providecommand \urlprefix  [0]{URL }%
\providecommand \Eprint [0]{\href }%
\providecommand \doibase [0]{https://doi.org/}%
\providecommand \selectlanguage [0]{\@gobble}%
\providecommand \bibinfo  [0]{\@secondoftwo}%
\providecommand \bibfield  [0]{\@secondoftwo}%
\providecommand \translation [1]{[#1]}%
\providecommand \BibitemOpen [0]{}%
\providecommand \bibitemStop [0]{}%
\providecommand \bibitemNoStop [0]{.\EOS\space}%
\providecommand \EOS [0]{\spacefactor3000\relax}%
\providecommand \BibitemShut  [1]{\csname bibitem#1\endcsname}%
\let\auto@bib@innerbib\@empty
\bibitem [{\citenamefont {Tanaka}\ \emph {et~al.}(2008)\citenamefont {Tanaka}, \citenamefont {Kontani}, \citenamefont {Naito}, \citenamefont {Naito}, \citenamefont {Hirashima}, \citenamefont {Yamada},\ and\ \citenamefont {Inoue}}]{tanaka_intrinsic_2008}%
  \BibitemOpen
  \bibfield  {author} {\bibinfo {author} {\bibfnamefont {T.}~\bibnamefont {Tanaka}}, \bibinfo {author} {\bibfnamefont {H.}~\bibnamefont {Kontani}}, \bibinfo {author} {\bibfnamefont {M.}~\bibnamefont {Naito}}, \bibinfo {author} {\bibfnamefont {T.}~\bibnamefont {Naito}}, \bibinfo {author} {\bibfnamefont {D.~S.}\ \bibnamefont {Hirashima}}, \bibinfo {author} {\bibfnamefont {K.}~\bibnamefont {Yamada}},\ and\ \bibinfo {author} {\bibfnamefont {J.}~\bibnamefont {Inoue}},\ }\bibfield  {title} {\bibinfo {title} {Intrinsic spin {Hall} effect and orbital {Hall} effect in $4d$ and $5d$ transition metals},\ }\href {https://doi.org/10.1103/PhysRevB.77.165117} {\bibfield  {journal} {\bibinfo  {journal} {Phys. Rev. B}\ }\textbf {\bibinfo {volume} {77}},\ \bibinfo {pages} {165117} (\bibinfo {year} {2008})}\BibitemShut {NoStop}%
\bibitem [{\citenamefont {Kontani}\ \emph {et~al.}(2009)\citenamefont {Kontani}, \citenamefont {Tanaka}, \citenamefont {Hirashima}, \citenamefont {Yamada},\ and\ \citenamefont {Inoue}}]{kontani_giant_2009}%
  \BibitemOpen
  \bibfield  {author} {\bibinfo {author} {\bibfnamefont {H.}~\bibnamefont {Kontani}}, \bibinfo {author} {\bibfnamefont {T.}~\bibnamefont {Tanaka}}, \bibinfo {author} {\bibfnamefont {D.~S.}\ \bibnamefont {Hirashima}}, \bibinfo {author} {\bibfnamefont {K.}~\bibnamefont {Yamada}},\ and\ \bibinfo {author} {\bibfnamefont {J.}~\bibnamefont {Inoue}},\ }\bibfield  {title} {\bibinfo {title} {Giant {Orbital} {Hall} {Effect} in {Transition} {Metals}: {Origin} of {Large} {Spin} and {Anomalous} {Hall} {Effects}},\ }\href {https://doi.org/10.1103/PhysRevLett.102.016601} {\bibfield  {journal} {\bibinfo  {journal} {Phys. Rev. Lett.}\ }\textbf {\bibinfo {volume} {102}},\ \bibinfo {pages} {016601} (\bibinfo {year} {2009})}\BibitemShut {NoStop}%
\bibitem [{\citenamefont {Go}\ \emph {et~al.}(2018)\citenamefont {Go}, \citenamefont {Jo}, \citenamefont {Kim},\ and\ \citenamefont {Lee}}]{go_intrinsic_2018}%
  \BibitemOpen
  \bibfield  {author} {\bibinfo {author} {\bibfnamefont {D.}~\bibnamefont {Go}}, \bibinfo {author} {\bibfnamefont {D.}~\bibnamefont {Jo}}, \bibinfo {author} {\bibfnamefont {C.}~\bibnamefont {Kim}},\ and\ \bibinfo {author} {\bibfnamefont {H.-W.}\ \bibnamefont {Lee}},\ }\bibfield  {title} {\bibinfo {title} {Intrinsic {Spin} and {Orbital} {Hall} {Effects} from {Orbital} {Texture}},\ }\href {https://doi.org/10.1103/PhysRevLett.121.086602} {\bibfield  {journal} {\bibinfo  {journal} {Phys. Rev. Lett.}\ }\textbf {\bibinfo {volume} {121}},\ \bibinfo {pages} {086602} (\bibinfo {year} {2018})}\BibitemShut {NoStop}%
\bibitem [{\citenamefont {Jo}\ \emph {et~al.}(2018)\citenamefont {Jo}, \citenamefont {Go},\ and\ \citenamefont {Lee}}]{jo_gigantic_2018}%
  \BibitemOpen
  \bibfield  {author} {\bibinfo {author} {\bibfnamefont {D.}~\bibnamefont {Jo}}, \bibinfo {author} {\bibfnamefont {D.}~\bibnamefont {Go}},\ and\ \bibinfo {author} {\bibfnamefont {H.-W.}\ \bibnamefont {Lee}},\ }\bibfield  {title} {\bibinfo {title} {Gigantic intrinsic orbital {Hall} effects in weakly spin-orbit coupled metals},\ }\href {https://doi.org/10.1103/PhysRevB.98.214405} {\bibfield  {journal} {\bibinfo  {journal} {Phys. Rev. B}\ }\textbf {\bibinfo {volume} {98}},\ \bibinfo {pages} {214405} (\bibinfo {year} {2018})}\BibitemShut {NoStop}%
\bibitem [{\citenamefont {Bhowal}\ and\ \citenamefont {Satpathy}(2020)}]{bhowal_intrinsic_2020}%
  \BibitemOpen
  \bibfield  {author} {\bibinfo {author} {\bibfnamefont {S.}~\bibnamefont {Bhowal}}\ and\ \bibinfo {author} {\bibfnamefont {S.}~\bibnamefont {Satpathy}},\ }\bibfield  {title} {\bibinfo {title} {Intrinsic orbital moment and prediction of a large orbital {Hall} effect in two-dimensional transition metal dichalcogenides},\ }\href {https://doi.org/10.1103/PhysRevB.101.121112} {\bibfield  {journal} {\bibinfo  {journal} {Phys. Rev. B}\ }\textbf {\bibinfo {volume} {101}},\ \bibinfo {pages} {121112} (\bibinfo {year} {2020})}\BibitemShut {NoStop}%
\bibitem [{\citenamefont {Ding}\ \emph {et~al.}(2020)\citenamefont {Ding}, \citenamefont {Ross}, \citenamefont {Go}, \citenamefont {Baldrati}, \citenamefont {Ren}, \citenamefont {Freimuth}, \citenamefont {Becker}, \citenamefont {Kammerbauer}, \citenamefont {Yang}, \citenamefont {Jakob}, \citenamefont {Mokrousov},\ and\ \citenamefont {Kl\"aui}}]{ding_harnessing_2020}%
  \BibitemOpen
  \bibfield  {author} {\bibinfo {author} {\bibfnamefont {S.}~\bibnamefont {Ding}}, \bibinfo {author} {\bibfnamefont {A.}~\bibnamefont {Ross}}, \bibinfo {author} {\bibfnamefont {D.}~\bibnamefont {Go}}, \bibinfo {author} {\bibfnamefont {L.}~\bibnamefont {Baldrati}}, \bibinfo {author} {\bibfnamefont {Z.}~\bibnamefont {Ren}}, \bibinfo {author} {\bibfnamefont {F.}~\bibnamefont {Freimuth}}, \bibinfo {author} {\bibfnamefont {S.}~\bibnamefont {Becker}}, \bibinfo {author} {\bibfnamefont {F.}~\bibnamefont {Kammerbauer}}, \bibinfo {author} {\bibfnamefont {J.}~\bibnamefont {Yang}}, \bibinfo {author} {\bibfnamefont {G.}~\bibnamefont {Jakob}}, \bibinfo {author} {\bibfnamefont {Y.}~\bibnamefont {Mokrousov}},\ and\ \bibinfo {author} {\bibfnamefont {M.}~\bibnamefont {Kl\"aui}},\ }\bibfield  {title} {\bibinfo {title} {Harnessing orbital-to-spin conversion of interfacial orbital currents for efficient spin-orbit torques},\ }\href {https://doi.org/10.1103/PhysRevLett.125.177201} {\bibfield  {journal} {\bibinfo  {journal} {Phys.
  Rev. Lett.}\ }\textbf {\bibinfo {volume} {125}},\ \bibinfo {pages} {177201} (\bibinfo {year} {2020})}\BibitemShut {NoStop}%
\bibitem [{\citenamefont {Kim}\ \emph {et~al.}(2021)\citenamefont {Kim}, \citenamefont {Go}, \citenamefont {Tsai}, \citenamefont {Jo}, \citenamefont {Kondou}, \citenamefont {Lee},\ and\ \citenamefont {Otani}}]{kim_nontrivial_2021}%
  \BibitemOpen
  \bibfield  {author} {\bibinfo {author} {\bibfnamefont {J.}~\bibnamefont {Kim}}, \bibinfo {author} {\bibfnamefont {D.}~\bibnamefont {Go}}, \bibinfo {author} {\bibfnamefont {H.}~\bibnamefont {Tsai}}, \bibinfo {author} {\bibfnamefont {D.}~\bibnamefont {Jo}}, \bibinfo {author} {\bibfnamefont {K.}~\bibnamefont {Kondou}}, \bibinfo {author} {\bibfnamefont {H.-W.}\ \bibnamefont {Lee}},\ and\ \bibinfo {author} {\bibfnamefont {Y.}~\bibnamefont {Otani}},\ }\bibfield  {title} {\bibinfo {title} {Nontrivial torque generation by orbital angular momentum injection in ferromagnetic-{metal/Cu/Al$_2$O$_3$} trilayers},\ }\href {https://doi.org/10.1103/PhysRevB.103.L020407} {\bibfield  {journal} {\bibinfo  {journal} {Phys. Rev. B}\ }\textbf {\bibinfo {volume} {103}},\ \bibinfo {pages} {L020407} (\bibinfo {year} {2021})}\BibitemShut {NoStop}%
\bibitem [{\citenamefont {Lee}\ \emph {et~al.}(2021{\natexlab{a}})\citenamefont {Lee}, \citenamefont {Kang}, \citenamefont {Go}, \citenamefont {Kim}, \citenamefont {Kang}, \citenamefont {Lee}, \citenamefont {Lee}, \citenamefont {Kang}, \citenamefont {Lee}, \citenamefont {Mokrousov}, \citenamefont {Kim}, \citenamefont {Kim}, \citenamefont {Lee},\ and\ \citenamefont {Park}}]{lee_efficient_2021}%
  \BibitemOpen
  \bibfield  {author} {\bibinfo {author} {\bibfnamefont {S.}~\bibnamefont {Lee}}, \bibinfo {author} {\bibfnamefont {M.-G.}\ \bibnamefont {Kang}}, \bibinfo {author} {\bibfnamefont {D.}~\bibnamefont {Go}}, \bibinfo {author} {\bibfnamefont {D.}~\bibnamefont {Kim}}, \bibinfo {author} {\bibfnamefont {J.-H.}\ \bibnamefont {Kang}}, \bibinfo {author} {\bibfnamefont {T.}~\bibnamefont {Lee}}, \bibinfo {author} {\bibfnamefont {G.-H.}\ \bibnamefont {Lee}}, \bibinfo {author} {\bibfnamefont {J.}~\bibnamefont {Kang}}, \bibinfo {author} {\bibfnamefont {N.~J.}\ \bibnamefont {Lee}}, \bibinfo {author} {\bibfnamefont {Y.}~\bibnamefont {Mokrousov}}, \bibinfo {author} {\bibfnamefont {S.}~\bibnamefont {Kim}}, \bibinfo {author} {\bibfnamefont {K.-J.}\ \bibnamefont {Kim}}, \bibinfo {author} {\bibfnamefont {K.-J.}\ \bibnamefont {Lee}},\ and\ \bibinfo {author} {\bibfnamefont {B.-G.}\ \bibnamefont {Park}},\ }\bibfield  {title} {\bibinfo {title} {Efficient conversion of orbital {Hall} current to spin current for spin-orbit torque
  switching},\ }\href {https://doi.org/10.1038/s42005-021-00737-7} {\bibfield  {journal} {\bibinfo  {journal} {Commun. Phys.}\ }\textbf {\bibinfo {volume} {4}},\ \bibinfo {pages} {234} (\bibinfo {year} {2021}{\natexlab{a}})}\BibitemShut {NoStop}%
\bibitem [{\citenamefont {Sala}\ and\ \citenamefont {Gambardella}(2022)}]{sala_giant_2022}%
  \BibitemOpen
  \bibfield  {author} {\bibinfo {author} {\bibfnamefont {G.}~\bibnamefont {Sala}}\ and\ \bibinfo {author} {\bibfnamefont {P.}~\bibnamefont {Gambardella}},\ }\bibfield  {title} {\bibinfo {title} {Giant orbital {Hall} effect and orbital-to-spin conversion in $3d$, $5d$, and $4f$ metallic heterostructures},\ }\href {https://doi.org/10.1103/PhysRevResearch.4.033037} {\bibfield  {journal} {\bibinfo  {journal} {Phys. Rev. Research}\ }\textbf {\bibinfo {volume} {4}},\ \bibinfo {pages} {033037} (\bibinfo {year} {2022})}\BibitemShut {NoStop}%
\bibitem [{\citenamefont {Hayashi}\ \emph {et~al.}(2023)\citenamefont {Hayashi}, \citenamefont {Jo}, \citenamefont {Go}, \citenamefont {Gao}, \citenamefont {Haku}, \citenamefont {Mokrousov},\ and\ \citenamefont {Ando}}]{hayashi_observation_2023}%
  \BibitemOpen
  \bibfield  {author} {\bibinfo {author} {\bibfnamefont {H.}~\bibnamefont {Hayashi}}, \bibinfo {author} {\bibfnamefont {D.}~\bibnamefont {Jo}}, \bibinfo {author} {\bibfnamefont {D.}~\bibnamefont {Go}}, \bibinfo {author} {\bibfnamefont {T.}~\bibnamefont {Gao}}, \bibinfo {author} {\bibfnamefont {S.}~\bibnamefont {Haku}}, \bibinfo {author} {\bibfnamefont {H.-W.}\ \bibnamefont {Mokrousov}, \bibfnamefont {Yuriy~Lee}},\ and\ \bibinfo {author} {\bibfnamefont {K.}~\bibnamefont {Ando}},\ }\bibfield  {title} {\bibinfo {title} {Observation of long-range orbital transport and giant orbital torque},\ }\href {https://doi.org/10.1038/s42005-023-01139-7} {\bibfield  {journal} {\bibinfo  {journal} {Commun. Phys.}\ }\textbf {\bibinfo {volume} {6}},\ \bibinfo {pages} {32} (\bibinfo {year} {2023})}\BibitemShut {NoStop}%
\bibitem [{\citenamefont {Lee}\ \emph {et~al.}(2021{\natexlab{b}})\citenamefont {Lee}, \citenamefont {Go}, \citenamefont {Park}, \citenamefont {Jeong}, \citenamefont {Ko}, \citenamefont {Yun}, \citenamefont {Jo}, \citenamefont {Lee}, \citenamefont {Go}, \citenamefont {Oh}, \citenamefont {Kim}, \citenamefont {Park}, \citenamefont {Min}, \citenamefont {Koo}, \citenamefont {Lee}, \citenamefont {Lee},\ and\ \citenamefont {Lee}}]{lee_orbital_2021}%
  \BibitemOpen
  \bibfield  {author} {\bibinfo {author} {\bibfnamefont {D.}~\bibnamefont {Lee}}, \bibinfo {author} {\bibfnamefont {D.}~\bibnamefont {Go}}, \bibinfo {author} {\bibfnamefont {H.-J.}\ \bibnamefont {Park}}, \bibinfo {author} {\bibfnamefont {W.}~\bibnamefont {Jeong}}, \bibinfo {author} {\bibfnamefont {H.-W.}\ \bibnamefont {Ko}}, \bibinfo {author} {\bibfnamefont {D.}~\bibnamefont {Yun}}, \bibinfo {author} {\bibfnamefont {D.}~\bibnamefont {Jo}}, \bibinfo {author} {\bibfnamefont {S.}~\bibnamefont {Lee}}, \bibinfo {author} {\bibfnamefont {G.}~\bibnamefont {Go}}, \bibinfo {author} {\bibfnamefont {J.~H.}\ \bibnamefont {Oh}}, \bibinfo {author} {\bibfnamefont {K.-J.}\ \bibnamefont {Kim}}, \bibinfo {author} {\bibfnamefont {B.-G.}\ \bibnamefont {Park}}, \bibinfo {author} {\bibfnamefont {B.-C.}\ \bibnamefont {Min}}, \bibinfo {author} {\bibfnamefont {H.~C.}\ \bibnamefont {Koo}}, \bibinfo {author} {\bibfnamefont {H.-W.}\ \bibnamefont {Lee}}, \bibinfo {author} {\bibfnamefont {O.}~\bibnamefont {Lee}},\ and\ \bibinfo {author}
  {\bibfnamefont {K.-J.}\ \bibnamefont {Lee}},\ }\bibfield  {title} {\bibinfo {title} {Orbital torque in magnetic bilayers},\ }\href {https://doi.org/10.1038/s41467-021-26650-9} {\bibfield  {journal} {\bibinfo  {journal} {Nat. Commun.}\ }\textbf {\bibinfo {volume} {12}},\ \bibinfo {pages} {6710} (\bibinfo {year} {2021}{\natexlab{b}})}\BibitemShut {NoStop}%
\bibitem [{\citenamefont {Choi}\ \emph {et~al.}(2023)\citenamefont {Choi}, \citenamefont {Jo}, \citenamefont {Ko}, \citenamefont {Go}, \citenamefont {Kim}, \citenamefont {Park}, \citenamefont {Kim}, \citenamefont {Min}, \citenamefont {Choi},\ and\ \citenamefont {Lee}}]{choi_observation_2023}%
  \BibitemOpen
  \bibfield  {author} {\bibinfo {author} {\bibfnamefont {Y.-G.}\ \bibnamefont {Choi}}, \bibinfo {author} {\bibfnamefont {D.}~\bibnamefont {Jo}}, \bibinfo {author} {\bibfnamefont {K.-H.}\ \bibnamefont {Ko}}, \bibinfo {author} {\bibfnamefont {D.}~\bibnamefont {Go}}, \bibinfo {author} {\bibfnamefont {K.-H.}\ \bibnamefont {Kim}}, \bibinfo {author} {\bibfnamefont {H.~G.}\ \bibnamefont {Park}}, \bibinfo {author} {\bibfnamefont {C.}~\bibnamefont {Kim}}, \bibinfo {author} {\bibfnamefont {B.-C.}\ \bibnamefont {Min}}, \bibinfo {author} {\bibfnamefont {G.-M.}\ \bibnamefont {Choi}},\ and\ \bibinfo {author} {\bibfnamefont {H.-W.}\ \bibnamefont {Lee}},\ }\bibfield  {title} {\bibinfo {title} {Observation of the orbital {Hall} effect in a light metal {Ti}},\ }\href {https://doi.org/10.1038/s41586-023-06101-9} {\bibfield  {journal} {\bibinfo  {journal} {Nature}\ }\textbf {\bibinfo {volume} {619}},\ \bibinfo {pages} {52} (\bibinfo {year} {2023})}\BibitemShut {NoStop}%
\bibitem [{\citenamefont {Lyalin}\ \emph {et~al.}(2023)\citenamefont {Lyalin}, \citenamefont {Alikhah}, \citenamefont {Berritta}, \citenamefont {Oppeneer},\ and\ \citenamefont {Kawakami}}]{lyalin_magneto-optical_2023}%
  \BibitemOpen
  \bibfield  {author} {\bibinfo {author} {\bibfnamefont {I.}~\bibnamefont {Lyalin}}, \bibinfo {author} {\bibfnamefont {S.}~\bibnamefont {Alikhah}}, \bibinfo {author} {\bibfnamefont {M.}~\bibnamefont {Berritta}}, \bibinfo {author} {\bibfnamefont {P.~M.}\ \bibnamefont {Oppeneer}},\ and\ \bibinfo {author} {\bibfnamefont {R.~K.}\ \bibnamefont {Kawakami}},\ }\bibfield  {title} {\bibinfo {title} {Magneto-optical detection of the orbital hall effect in chromium},\ }\href {https://doi.org/10.1103/PhysRevLett.131.156702} {\bibfield  {journal} {\bibinfo  {journal} {Phys. Rev. Lett.}\ }\textbf {\bibinfo {volume} {131}},\ \bibinfo {pages} {156702} (\bibinfo {year} {2023})}\BibitemShut {NoStop}%
\bibitem [{\citenamefont {Go}\ and\ \citenamefont {Lee}(2020)}]{go_orbital_2020}%
  \BibitemOpen
  \bibfield  {author} {\bibinfo {author} {\bibfnamefont {D.}~\bibnamefont {Go}}\ and\ \bibinfo {author} {\bibfnamefont {H.-W.}\ \bibnamefont {Lee}},\ }\bibfield  {title} {\bibinfo {title} {Orbital torque: {Torque} generation by orbital current injection},\ }\href {https://doi.org/10.1103/PhysRevResearch.2.013177} {\bibfield  {journal} {\bibinfo  {journal} {Phys. Rev. Res.}\ }\textbf {\bibinfo {volume} {2}},\ \bibinfo {pages} {013177} (\bibinfo {year} {2020})}\BibitemShut {NoStop}%
\bibitem [{\citenamefont {Fan}\ \emph {et~al.}(2014)\citenamefont {Fan}, \citenamefont {Celik}, \citenamefont {Wu}, \citenamefont {Ni}, \citenamefont {Lee}, \citenamefont {Lorenz},\ and\ \citenamefont {Xiao}}]{fan_quantifying_2014}%
  \BibitemOpen
  \bibfield  {author} {\bibinfo {author} {\bibfnamefont {X.}~\bibnamefont {Fan}}, \bibinfo {author} {\bibfnamefont {H.}~\bibnamefont {Celik}}, \bibinfo {author} {\bibfnamefont {J.}~\bibnamefont {Wu}}, \bibinfo {author} {\bibfnamefont {C.}~\bibnamefont {Ni}}, \bibinfo {author} {\bibfnamefont {K.-J.}\ \bibnamefont {Lee}}, \bibinfo {author} {\bibfnamefont {V.~O.}\ \bibnamefont {Lorenz}},\ and\ \bibinfo {author} {\bibfnamefont {J.~Q.}\ \bibnamefont {Xiao}},\ }\bibfield  {title} {\bibinfo {title} {Quantifying interface and bulk contributions to spin–orbit torque in magnetic bilayers},\ }\href {https://doi.org/10.1038/ncomms4042} {\bibfield  {journal} {\bibinfo  {journal} {Nat. Commun.}\ }\textbf {\bibinfo {volume} {5}},\ \bibinfo {pages} {3042} (\bibinfo {year} {2014})}\BibitemShut {NoStop}%
\bibitem [{\citenamefont {Fan}\ \emph {et~al.}(2016)\citenamefont {Fan}, \citenamefont {Mellnik}, \citenamefont {Wang}, \citenamefont {Reynolds}, \citenamefont {Wang}, \citenamefont {Celik}, \citenamefont {Lorenz}, \citenamefont {Ralph},\ and\ \citenamefont {Xiao}}]{fan_all-optical_2016}%
  \BibitemOpen
  \bibfield  {author} {\bibinfo {author} {\bibfnamefont {X.}~\bibnamefont {Fan}}, \bibinfo {author} {\bibfnamefont {A.~R.}\ \bibnamefont {Mellnik}}, \bibinfo {author} {\bibfnamefont {W.}~\bibnamefont {Wang}}, \bibinfo {author} {\bibfnamefont {N.}~\bibnamefont {Reynolds}}, \bibinfo {author} {\bibfnamefont {T.}~\bibnamefont {Wang}}, \bibinfo {author} {\bibfnamefont {H.}~\bibnamefont {Celik}}, \bibinfo {author} {\bibfnamefont {V.~O.}\ \bibnamefont {Lorenz}}, \bibinfo {author} {\bibfnamefont {D.~C.}\ \bibnamefont {Ralph}},\ and\ \bibinfo {author} {\bibfnamefont {J.~Q.}\ \bibnamefont {Xiao}},\ }\bibfield  {title} {\bibinfo {title} {All-optical vector measurement of spin-orbit-induced torques using both polar and quadratic magneto-optic {Kerr} effects},\ }\href {https://doi.org/10.1063/1.4962402} {\bibfield  {journal} {\bibinfo  {journal} {Applied Physics Letters}\ }\textbf {\bibinfo {volume} {109}},\ \bibinfo {pages} {122406} (\bibinfo {year} {2016})}\BibitemShut {NoStop}%
\bibitem [{\citenamefont {Nguyen}\ \emph {et~al.}(2016)\citenamefont {Nguyen}, \citenamefont {Ralph},\ and\ \citenamefont {Buhrman}}]{nguyen_spin_2016}%
  \BibitemOpen
  \bibfield  {author} {\bibinfo {author} {\bibfnamefont {M.-H.}\ \bibnamefont {Nguyen}}, \bibinfo {author} {\bibfnamefont {D.~C.}\ \bibnamefont {Ralph}},\ and\ \bibinfo {author} {\bibfnamefont {R.~A.}\ \bibnamefont {Buhrman}},\ }\bibfield  {title} {\bibinfo {title} {Spin torque study of the spin {Hall} conductivity and spin diffusion length in platinum thin films with varying resistivity},\ }\href {https://doi.org/10.1103/PhysRevLett.116.126601} {\bibfield  {journal} {\bibinfo  {journal} {Phys. Rev. Lett.}\ }\textbf {\bibinfo {volume} {116}},\ \bibinfo {pages} {126601} (\bibinfo {year} {2016})}\BibitemShut {NoStop}%
\bibitem [{\citenamefont {Stamm}\ \emph {et~al.}(2017)\citenamefont {Stamm}, \citenamefont {Murer}, \citenamefont {Berritta}, \citenamefont {Feng}, \citenamefont {Gabureac}, \citenamefont {Oppeneer},\ and\ \citenamefont {Gambardella}}]{stamm_magneto-optical_2017}%
  \BibitemOpen
  \bibfield  {author} {\bibinfo {author} {\bibfnamefont {C.}~\bibnamefont {Stamm}}, \bibinfo {author} {\bibfnamefont {C.}~\bibnamefont {Murer}}, \bibinfo {author} {\bibfnamefont {M.}~\bibnamefont {Berritta}}, \bibinfo {author} {\bibfnamefont {J.}~\bibnamefont {Feng}}, \bibinfo {author} {\bibfnamefont {M.}~\bibnamefont {Gabureac}}, \bibinfo {author} {\bibfnamefont {P.~M.}\ \bibnamefont {Oppeneer}},\ and\ \bibinfo {author} {\bibfnamefont {P.}~\bibnamefont {Gambardella}},\ }\bibfield  {title} {\bibinfo {title} {Magneto-{Optical} {Detection} of the {Spin} {Hall} {Effect} in {Pt} and {W} {Thin} {Films}},\ }\href {https://doi.org/10.1103/PhysRevLett.119.087203} {\bibfield  {journal} {\bibinfo  {journal} {Phys. Rev. Lett.}\ }\textbf {\bibinfo {volume} {119}},\ \bibinfo {pages} {087203} (\bibinfo {year} {2017})}\BibitemShut {NoStop}%
\bibitem [{\citenamefont {Bass}\ and\ \citenamefont {Pratt}(2007)}]{bass_spin-diffusion_2007}%
  \BibitemOpen
  \bibfield  {author} {\bibinfo {author} {\bibfnamefont {J.}~\bibnamefont {Bass}}\ and\ \bibinfo {author} {\bibfnamefont {W.~P.}\ \bibnamefont {Pratt}},\ }\bibfield  {title} {\bibinfo {title} {Spin-diffusion lengths in metals and alloys, and spin-flipping at metal/metal interfaces: an experimentalist’s critical review},\ }\href {https://doi.org/10.1088/0953-8984/19/18/183201} {\bibfield  {journal} {\bibinfo  {journal} {J. Phys.: Condens. Matter}\ }\textbf {\bibinfo {volume} {19}},\ \bibinfo {pages} {183201} (\bibinfo {year} {2007})}\BibitemShut {NoStop}%
\bibitem [{\citenamefont {Zhang}\ \emph {et~al.}(2015)\citenamefont {Zhang}, \citenamefont {Han}, \citenamefont {Jiang}, \citenamefont {Yang},\ and\ \citenamefont {S.~P.~Parkin}}]{zhang_role_2015}%
  \BibitemOpen
  \bibfield  {author} {\bibinfo {author} {\bibfnamefont {W.}~\bibnamefont {Zhang}}, \bibinfo {author} {\bibfnamefont {W.}~\bibnamefont {Han}}, \bibinfo {author} {\bibfnamefont {X.}~\bibnamefont {Jiang}}, \bibinfo {author} {\bibfnamefont {S.-H.}\ \bibnamefont {Yang}},\ and\ \bibinfo {author} {\bibfnamefont {S.}~\bibnamefont {S.~P.~Parkin}},\ }\bibfield  {title} {\bibinfo {title} {Role of transparency of platinum–ferromagnet interfaces in determining the intrinsic magnitude of the spin {Hall} effect},\ }\href {https://doi.org/10.1038/nphys3304} {\bibfield  {journal} {\bibinfo  {journal} {Nature Physics}\ }\textbf {\bibinfo {volume} {11}},\ \bibinfo {pages} {496} (\bibinfo {year} {2015})}\BibitemShut {NoStop}%
\bibitem [{\citenamefont {Salemi}\ and\ \citenamefont {Oppeneer}(2022)}]{salemi_first-principles_2022}%
  \BibitemOpen
  \bibfield  {author} {\bibinfo {author} {\bibfnamefont {L.}~\bibnamefont {Salemi}}\ and\ \bibinfo {author} {\bibfnamefont {P.~M.}\ \bibnamefont {Oppeneer}},\ }\bibfield  {title} {\bibinfo {title} {First-principles theory of intrinsic spin and orbital {Hall} and {Nernst} effects in metallic monoatomic crystals},\ }\href {https://doi.org/10.1103/PhysRevMaterials.6.095001} {\bibfield  {journal} {\bibinfo  {journal} {Phys. Rev. Mater.}\ }\textbf {\bibinfo {volume} {6}},\ \bibinfo {pages} {095001} (\bibinfo {year} {2022})}\BibitemShut {NoStop}%
\bibitem [{\citenamefont {Wang}\ \emph {et~al.}(2019{\natexlab{a}})\citenamefont {Wang}, \citenamefont {Wang}, \citenamefont {Amin}, \citenamefont {Wang}, \citenamefont {Radhakrishnan}, \citenamefont {Davidson}, \citenamefont {Allen}, \citenamefont {Silva}, \citenamefont {Ohldag}, \citenamefont {Balzar}, \citenamefont {Zink}, \citenamefont {Haney}, \citenamefont {Xiao}, \citenamefont {Cahill}, \citenamefont {Lorenz},\ and\ \citenamefont {Fan}}]{wang_anomalous_2019}%
  \BibitemOpen
  \bibfield  {author} {\bibinfo {author} {\bibfnamefont {W.}~\bibnamefont {Wang}}, \bibinfo {author} {\bibfnamefont {T.}~\bibnamefont {Wang}}, \bibinfo {author} {\bibfnamefont {V.~P.}\ \bibnamefont {Amin}}, \bibinfo {author} {\bibfnamefont {Y.}~\bibnamefont {Wang}}, \bibinfo {author} {\bibfnamefont {A.}~\bibnamefont {Radhakrishnan}}, \bibinfo {author} {\bibfnamefont {A.}~\bibnamefont {Davidson}}, \bibinfo {author} {\bibfnamefont {S.~R.}\ \bibnamefont {Allen}}, \bibinfo {author} {\bibfnamefont {T.~J.}\ \bibnamefont {Silva}}, \bibinfo {author} {\bibfnamefont {H.}~\bibnamefont {Ohldag}}, \bibinfo {author} {\bibfnamefont {D.}~\bibnamefont {Balzar}}, \bibinfo {author} {\bibfnamefont {B.~L.}\ \bibnamefont {Zink}}, \bibinfo {author} {\bibfnamefont {P.~M.}\ \bibnamefont {Haney}}, \bibinfo {author} {\bibfnamefont {J.~Q.}\ \bibnamefont {Xiao}}, \bibinfo {author} {\bibfnamefont {D.~G.}\ \bibnamefont {Cahill}}, \bibinfo {author} {\bibfnamefont {V.~O.}\ \bibnamefont {Lorenz}},\ and\ \bibinfo {author} {\bibfnamefont
  {X.}~\bibnamefont {Fan}},\ }\bibfield  {title} {\bibinfo {title} {Anomalous spin–orbit torques in magnetic single-layer films},\ }\href {https://doi.org/10.1038/s41565-019-0504-0} {\bibfield  {journal} {\bibinfo  {journal} {Nat. Nanotechnol.}\ }\textbf {\bibinfo {volume} {14}},\ \bibinfo {pages} {819} (\bibinfo {year} {2019}{\natexlab{a}})}\BibitemShut {NoStop}%
\bibitem [{\citenamefont {Go}\ \emph {et~al.}(2023{\natexlab{a}})\citenamefont {Go}, \citenamefont {Jo}, \citenamefont {Kim}, \citenamefont {Lee}, \citenamefont {Kang}, \citenamefont {Park}, \citenamefont {Bl\"ugel}, \citenamefont {Lee},\ and\ \citenamefont {Mokrousov}}]{go_long-range_2023}%
  \BibitemOpen
  \bibfield  {author} {\bibinfo {author} {\bibfnamefont {D.}~\bibnamefont {Go}}, \bibinfo {author} {\bibfnamefont {D.}~\bibnamefont {Jo}}, \bibinfo {author} {\bibfnamefont {K.-W.}\ \bibnamefont {Kim}}, \bibinfo {author} {\bibfnamefont {S.}~\bibnamefont {Lee}}, \bibinfo {author} {\bibfnamefont {M.-G.}\ \bibnamefont {Kang}}, \bibinfo {author} {\bibfnamefont {B.-G.}\ \bibnamefont {Park}}, \bibinfo {author} {\bibfnamefont {S.}~\bibnamefont {Bl\"ugel}}, \bibinfo {author} {\bibfnamefont {H.-W.}\ \bibnamefont {Lee}},\ and\ \bibinfo {author} {\bibfnamefont {Y.}~\bibnamefont {Mokrousov}},\ }\bibfield  {title} {\bibinfo {title} {Long-range orbital torque by momentum-space hotspots},\ }\href {https://doi.org/10.1103/PhysRevLett.130.246701} {\bibfield  {journal} {\bibinfo  {journal} {Phys. Rev. Lett.}\ }\textbf {\bibinfo {volume} {130}},\ \bibinfo {pages} {246701} (\bibinfo {year} {2023}{\natexlab{a}})}\BibitemShut {NoStop}%
\bibitem [{\citenamefont {Wang}\ \emph {et~al.}(2019{\natexlab{b}})\citenamefont {Wang}, \citenamefont {Zhu}, \citenamefont {Yang}, \citenamefont {Lee}, \citenamefont {Mishra}, \citenamefont {Go}, \citenamefont {Oh}, \citenamefont {Kim}, \citenamefont {Cai}, \citenamefont {Liu}, \citenamefont {Pollard}, \citenamefont {Shi}, \citenamefont {Lee}, \citenamefont {Teo}, \citenamefont {Wu}, \citenamefont {Lee},\ and\ \citenamefont {Yang}}]{wang_magnetization_2019}%
  \BibitemOpen
  \bibfield  {author} {\bibinfo {author} {\bibfnamefont {Y.}~\bibnamefont {Wang}}, \bibinfo {author} {\bibfnamefont {D.}~\bibnamefont {Zhu}}, \bibinfo {author} {\bibfnamefont {Y.}~\bibnamefont {Yang}}, \bibinfo {author} {\bibfnamefont {K.}~\bibnamefont {Lee}}, \bibinfo {author} {\bibfnamefont {R.}~\bibnamefont {Mishra}}, \bibinfo {author} {\bibfnamefont {G.}~\bibnamefont {Go}}, \bibinfo {author} {\bibfnamefont {S.-H.}\ \bibnamefont {Oh}}, \bibinfo {author} {\bibfnamefont {D.-H.}\ \bibnamefont {Kim}}, \bibinfo {author} {\bibfnamefont {K.}~\bibnamefont {Cai}}, \bibinfo {author} {\bibfnamefont {E.}~\bibnamefont {Liu}}, \bibinfo {author} {\bibfnamefont {S.~D.}\ \bibnamefont {Pollard}}, \bibinfo {author} {\bibfnamefont {S.}~\bibnamefont {Shi}}, \bibinfo {author} {\bibfnamefont {J.}~\bibnamefont {Lee}}, \bibinfo {author} {\bibfnamefont {K.~L.}\ \bibnamefont {Teo}}, \bibinfo {author} {\bibfnamefont {Y.}~\bibnamefont {Wu}}, \bibinfo {author} {\bibfnamefont {K.-J.}\ \bibnamefont {Lee}},\ and\ \bibinfo {author}
  {\bibfnamefont {H.}~\bibnamefont {Yang}},\ }\bibfield  {title} {\bibinfo {title} {Magnetization switching by magnon-mediated spin torque through an antiferromagnetic insulator},\ }\href {https://doi.org/10.1126/science.aav8076} {\bibfield  {journal} {\bibinfo  {journal} {Science}\ }\textbf {\bibinfo {volume} {366}},\ \bibinfo {pages} {1125} (\bibinfo {year} {2019}{\natexlab{b}})}\BibitemShut {NoStop}%
\bibitem [{\citenamefont {Zhu}\ \emph {et~al.}(2021{\natexlab{a}})\citenamefont {Zhu}, \citenamefont {Zhu},\ and\ \citenamefont {Buhrman}}]{zhu_fully_2021}%
  \BibitemOpen
  \bibfield  {author} {\bibinfo {author} {\bibfnamefont {L.}~\bibnamefont {Zhu}}, \bibinfo {author} {\bibfnamefont {L.}~\bibnamefont {Zhu}},\ and\ \bibinfo {author} {\bibfnamefont {R.~A.}\ \bibnamefont {Buhrman}},\ }\bibfield  {title} {\bibinfo {title} {Fully spin-transparent magnetic interfaces enabled by the insertion of a thin paramagnetic {NiO} layer},\ }\href {https://doi.org/10.1103/PhysRevLett.126.107204} {\bibfield  {journal} {\bibinfo  {journal} {Phys. Rev. Lett.}\ }\textbf {\bibinfo {volume} {126}},\ \bibinfo {pages} {107204} (\bibinfo {year} {2021}{\natexlab{a}})}\BibitemShut {NoStop}%
\bibitem [{\citenamefont {Zhu}\ \emph {et~al.}(2021{\natexlab{b}})\citenamefont {Zhu}, \citenamefont {Ralph},\ and\ \citenamefont {Buhrman}}]{zhu_maximizing_2021}%
  \BibitemOpen
  \bibfield  {author} {\bibinfo {author} {\bibfnamefont {L.}~\bibnamefont {Zhu}}, \bibinfo {author} {\bibfnamefont {D.~C.}\ \bibnamefont {Ralph}},\ and\ \bibinfo {author} {\bibfnamefont {R.~A.}\ \bibnamefont {Buhrman}},\ }\bibfield  {title} {\bibinfo {title} {{Maximizing spin-orbit torque generated by the spin {Hall} effect of Pt}},\ }\href {https://doi.org/10.1063/5.0059171} {\bibfield  {journal} {\bibinfo  {journal} {Appl. Phys. Rev.}\ }\textbf {\bibinfo {volume} {8}} (\bibinfo {year} {2021}{\natexlab{b}})}\BibitemShut {NoStop}%
\bibitem [{\citenamefont {Zhu}\ \emph {et~al.}(2022)\citenamefont {Zhu}, \citenamefont {Zhang}, \citenamefont {Fu}, \citenamefont {Hao}, \citenamefont {Hamzi\ifmmode~\acute{c}\else \'{c}\fi{}}, \citenamefont {Yang}, \citenamefont {Zhang}, \citenamefont {Zhang}, \citenamefont {Du}, \citenamefont {Xiong}, \citenamefont {Shi}, \citenamefont {Yan}, \citenamefont {Zhang}, \citenamefont {Fert},\ and\ \citenamefont {Zhao}}]{zhu_sign_2022}%
  \BibitemOpen
  \bibfield  {author} {\bibinfo {author} {\bibfnamefont {D.}~\bibnamefont {Zhu}}, \bibinfo {author} {\bibfnamefont {T.}~\bibnamefont {Zhang}}, \bibinfo {author} {\bibfnamefont {X.}~\bibnamefont {Fu}}, \bibinfo {author} {\bibfnamefont {R.}~\bibnamefont {Hao}}, \bibinfo {author} {\bibfnamefont {A.}~\bibnamefont {Hamzi\ifmmode~\acute{c}\else \'{c}\fi{}}}, \bibinfo {author} {\bibfnamefont {H.}~\bibnamefont {Yang}}, \bibinfo {author} {\bibfnamefont {X.}~\bibnamefont {Zhang}}, \bibinfo {author} {\bibfnamefont {H.}~\bibnamefont {Zhang}}, \bibinfo {author} {\bibfnamefont {A.}~\bibnamefont {Du}}, \bibinfo {author} {\bibfnamefont {D.}~\bibnamefont {Xiong}}, \bibinfo {author} {\bibfnamefont {K.}~\bibnamefont {Shi}}, \bibinfo {author} {\bibfnamefont {S.}~\bibnamefont {Yan}}, \bibinfo {author} {\bibfnamefont {S.}~\bibnamefont {Zhang}}, \bibinfo {author} {\bibfnamefont {A.}~\bibnamefont {Fert}},\ and\ \bibinfo {author} {\bibfnamefont {W.}~\bibnamefont {Zhao}},\ }\bibfield  {title} {\bibinfo {title} {Sign change of
  spin-orbit torque in $\mathrm{Pt}/\mathrm{NiO}/\mathrm{CoFeB}$ structures},\ }\href {https://doi.org/10.1103/PhysRevLett.128.217702} {\bibfield  {journal} {\bibinfo  {journal} {Phys. Rev. Lett.}\ }\textbf {\bibinfo {volume} {128}},\ \bibinfo {pages} {217702} (\bibinfo {year} {2022})}\BibitemShut {NoStop}%
\bibitem [{\citenamefont {Holloway}(1981)}]{holloway_chemisorption_1981}%
  \BibitemOpen
  \bibfield  {author} {\bibinfo {author} {\bibfnamefont {P.~H.}\ \bibnamefont {Holloway}},\ }\bibfield  {title} {\bibinfo {title} {{Chemisorption and oxide formation on metals: Oxygen–nickel reaction}},\ }\href {https://doi.org/10.1116/1.570847} {\bibfield  {journal} {\bibinfo  {journal} {Journal of Vacuum Science and Technology}\ }\textbf {\bibinfo {volume} {18}},\ \bibinfo {pages} {653} (\bibinfo {year} {1981})}\BibitemShut {NoStop}%
\bibitem [{\citenamefont {Mitchell}(1982)}]{mitchell_kinetic_1982}%
  \BibitemOpen
  \bibfield  {author} {\bibinfo {author} {\bibfnamefont {D.}~\bibnamefont {Mitchell}},\ }\bibfield  {title} {\bibinfo {title} {A kinetic study of the initial oxidation of {Ni} (111) and (211) surfaces by {RHEED} and {X}-ray emission},\ }\href {https://doi.org/10.1016/0039-6028(82)90704-X} {\bibfield  {journal} {\bibinfo  {journal} {Surface science}\ }\textbf {\bibinfo {volume} {114}},\ \bibinfo {pages} {546} (\bibinfo {year} {1982})}\BibitemShut {NoStop}%
\bibitem [{\citenamefont {Lambers}\ \emph {et~al.}(1996)\citenamefont {Lambers}, \citenamefont {Dykstal}, \citenamefont {Seo}, \citenamefont {Rowe},\ and\ \citenamefont {Holloway}}]{lambers_room-temerature_1996}%
  \BibitemOpen
  \bibfield  {author} {\bibinfo {author} {\bibfnamefont {E.~S.}\ \bibnamefont {Lambers}}, \bibinfo {author} {\bibfnamefont {C.~N.}\ \bibnamefont {Dykstal}}, \bibinfo {author} {\bibfnamefont {J.~M.}\ \bibnamefont {Seo}}, \bibinfo {author} {\bibfnamefont {J.~E.}\ \bibnamefont {Rowe}},\ and\ \bibinfo {author} {\bibfnamefont {P.~H.}\ \bibnamefont {Holloway}},\ }\bibfield  {title} {\bibinfo {title} {Room-temerature oxidation of {Ni}(110) at low and atmospheric oxygen pressures},\ }\href {https://doi.org/10.1007/BF01046987} {\bibfield  {journal} {\bibinfo  {journal} {Oxidation of Metals}\ }\textbf {\bibinfo {volume} {45}},\ \bibinfo {pages} {301} (\bibinfo {year} {1996})}\BibitemShut {NoStop}%
\bibitem [{\citenamefont {Kitakatsu}\ \emph {et~al.}(1998)\citenamefont {Kitakatsu}, \citenamefont {Maurice}, \citenamefont {Hinnen},\ and\ \citenamefont {Marcus}}]{kitakatsu_surface_1998}%
  \BibitemOpen
  \bibfield  {author} {\bibinfo {author} {\bibfnamefont {N.}~\bibnamefont {Kitakatsu}}, \bibinfo {author} {\bibfnamefont {V.}~\bibnamefont {Maurice}}, \bibinfo {author} {\bibfnamefont {C.}~\bibnamefont {Hinnen}},\ and\ \bibinfo {author} {\bibfnamefont {P.}~\bibnamefont {Marcus}},\ }\bibfield  {title} {\bibinfo {title} {{Surface hydroxylation and local structure of NiO thin films formed on Ni(111)}},\ }\href {https://doi.org/10.1016/s0039-6028(98)00089-2} {\bibfield  {journal} {\bibinfo  {journal} {Surface Science}\ }\textbf {\bibinfo {volume} {407}},\ \bibinfo {pages} {36} (\bibinfo {year} {1998})}\BibitemShut {NoStop}%
\bibitem [{\citenamefont {Neumann}\ \emph {et~al.}(2020)\citenamefont {Neumann}, \citenamefont {Mook}, \citenamefont {Henk},\ and\ \citenamefont {Mertig}}]{neumann_orbital_2020}%
  \BibitemOpen
  \bibfield  {author} {\bibinfo {author} {\bibfnamefont {R.~R.}\ \bibnamefont {Neumann}}, \bibinfo {author} {\bibfnamefont {A.}~\bibnamefont {Mook}}, \bibinfo {author} {\bibfnamefont {J.}~\bibnamefont {Henk}},\ and\ \bibinfo {author} {\bibfnamefont {I.}~\bibnamefont {Mertig}},\ }\bibfield  {title} {\bibinfo {title} {Orbital magnetic moment of magnons},\ }\href {https://doi.org/10.1103/PhysRevLett.125.117209} {\bibfield  {journal} {\bibinfo  {journal} {Phys. Rev. Lett.}\ }\textbf {\bibinfo {volume} {125}},\ \bibinfo {pages} {117209} (\bibinfo {year} {2020})}\BibitemShut {NoStop}%
\bibitem [{\citenamefont {Fishman}\ \emph {et~al.}(2022)\citenamefont {Fishman}, \citenamefont {Gardner},\ and\ \citenamefont {Okamoto}}]{fishman_orbital_2023}%
  \BibitemOpen
  \bibfield  {author} {\bibinfo {author} {\bibfnamefont {R.~S.}\ \bibnamefont {Fishman}}, \bibinfo {author} {\bibfnamefont {J.~S.}\ \bibnamefont {Gardner}},\ and\ \bibinfo {author} {\bibfnamefont {S.}~\bibnamefont {Okamoto}},\ }\bibfield  {title} {\bibinfo {title} {Orbital angular momentum of magnons in collinear magnets},\ }\href {https://doi.org/10.1103/PhysRevLett.129.167202} {\bibfield  {journal} {\bibinfo  {journal} {Phys. Rev. Lett.}\ }\textbf {\bibinfo {volume} {129}},\ \bibinfo {pages} {167202} (\bibinfo {year} {2022})}\BibitemShut {NoStop}%
\bibitem [{\citenamefont {Go}\ \emph {et~al.}(2023{\natexlab{b}})\citenamefont {Go}, \citenamefont {An}, \citenamefont {Lee},\ and\ \citenamefont {Kim}}]{go_intrinsic_2023}%
  \BibitemOpen
  \bibfield  {author} {\bibinfo {author} {\bibfnamefont {G.}~\bibnamefont {Go}}, \bibinfo {author} {\bibfnamefont {D.}~\bibnamefont {An}}, \bibinfo {author} {\bibfnamefont {H.-W.}\ \bibnamefont {Lee}},\ and\ \bibinfo {author} {\bibfnamefont {S.~K.}\ \bibnamefont {Kim}},\ }\href@noop {} {\bibinfo {title} {Intrinsic magnon orbital {Hall} effect in honeycomb antiferromagnets}} (\bibinfo {year} {2023}{\natexlab{b}}),\ \Eprint {https://arxiv.org/abs/2303.11687} {arXiv:2303.11687 [cond-mat.mes-hall]} \BibitemShut {NoStop}%
\end{thebibliography}%

\end{document}